# Metamemory: Exploring the Resilience of Older Internal Migrants

Xiaoxiao Wang[1], Jingjing Zhang[1], Huize Wan[1], Weiwei Zhang[2] and Yuan Yao[1,*]

[1]*Beijing Jiaotong University, School of Architecture and Design, Beijing, China*
[2]*Beijing University of Posts and Telecommunications, School of Digital Media & Design Arts, Beijing, China*

#### Abstract
Immigration and aging have always been significant topics of discussion in society, concerning the stability and future development of a country and its people. Research in the field of HCI on immigration and aging has primarily focused on their practical needs but has paid less attention to the adaptability issues of older internal migrants moving with their families. In this study, we investigate the challenges older internal migrants face in adapting socially, using metadata surveys and semi-structured interviews to delve into their life struggles and resilience sources. Our findings highlight the older internal migrants' remarkable resilience, particularly evident in their reminiscences. We explore the integration of reminiscences with the metaverse, identifying the necessary conditions to create a "Metamemory". We introduce a novel design for a metaverse scene that bridges past and present experiences. This aims to encourage discussions on enhancing older internal migrants' reminiscence, leveraging the metaverse's positive potential, and devising strategies to more effectively address older internal migrants' concerns in the future.

#### Keywords
Older Internal Migrants, Metaverse, Resilience, Memory

## 1. Introduction

Recent years have highlighted HCI's focus on the adaptation and welfare of new immigrants, emphasizing support from design, technological, and cultural aspects. This initiative seeks to facilitate their swift adjustment to new socio-cultural contexts. Among the diverse immigrant groups, the number of older internal migrants is on the rise. While some choose to relocate to more habitable cities post-retirement [1, 2], others move passively alongside their families [3, 4, 5]. Despite being within the same country, different regions exhibit distinct living habits and cultural identities. Hence, we also consider the study of immigrants to better understand the challenges older internal migrants face. These individuals typically display independence and a strong sense of identity but must adjust to new living and social settings. They encounter issues like shrinking social networks, reduced participation in social activities, and declining everyday skills. Relocating to new cities, they might struggle to blend into new communities and encounter dialect obstacles, which can create social difficulties (Figure 1). The changes in life pace brought about by migration have a profound effect on older adults.

Furthermore, some studies indicate that therapies involving cultural identity, memories, and group recollection can help new immigrant groups gain resilience [6], reduce feelings of alienation, and acquire a sense of unity from an individual subjective perspective [7, 8]. At the same time, with the continuous development of the metaverse concept and technology, an increasing amount of work has focused on the design potential of VR for older adults. Common applications include using VR technology in the daily lives of older adults, such as aiding in medical care and designing age-friendly systems [9, 10, 11], VR simulations for wayfinding [12], and stress relief [13]. They also develop immersive content mechanisms to enhance social connections among family and friends [14, 15]. Many studies also examine the impact of immersive digital tools on older adults, investigating whether they contribute to digital inclusion or exacerbate the digital divide [16, 17]. For example, social VR is utilized to challenge aging stereotypes and promote healthy aging [18]. Design of VR game difficulty levels tailored for older adults [19]. A wealth of previous research provides potential technical and design support for our subsequent research.

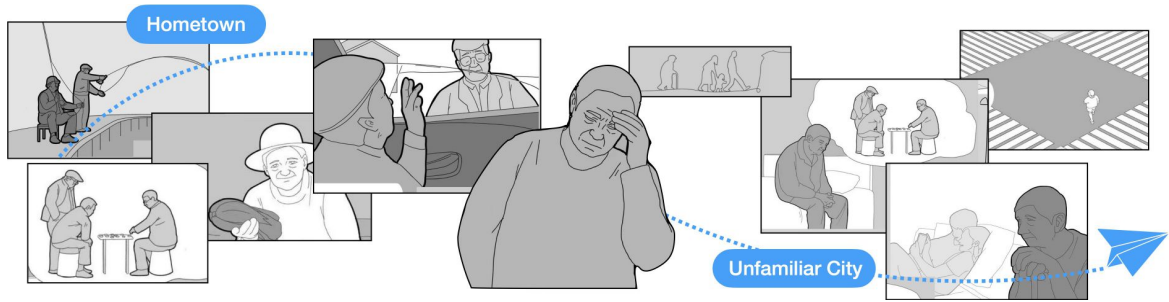

**Figure 1:** Like migratory birds, some older migrants move from their familiar hometowns to unfamiliar living places for their children, embarking on a new round of dedication. Despite reuniting with their children, they still feel like outsiders.

All this work confirms the benefits of reminiscence and VR technology for migrants, yet emphasizes the necessity of field surveys focused on the specific challenges and needs of older internal migrants. Engaging 9 participants, we conducted in-depth semi-structured interviews and discussions on reminiscence design. Findings from the literature review and interviews indicate a significant positive impact of immersive VR tools on these older internal migrants. This insight prompts us to explore the metaverse's potential as a platform for reminiscence and healing further. It appears crucial for older internal migrants to reconnect with their previous lives, deriving courage and strength from their past experiences to pursue new aspirations. Consequently, we advocate for the adoption of metaverse environments as platforms for reminiscence. This approach aims to facilitate the integration of their former cultural identities and social experiences into their future endeavors, ultimately reigniting their passion for life and resilience.

## 2. About Older Internal Migration

To gain insight into the challenges and needs of older internal migrants, we conducted an internet-based metadata survey on "older internal migrants". According to the 2020 China Migration Population Development Report, the country's migrant population is now largely moving as families, with the count nearing 18 million—about double the population of New York. Of these migrants, 43% relocated to care for grandchildren, and 25% moved to reunite with children or for retirement in a new area. The report "Internet Social Support for New Urban Immigrants under the Context of New Urbanization" [1], focusing on urban residents over 60 in parts of Jiangsu, China, showed significant demand for information support among older migrants, at 61.1%. The need for emotional support and companionship was also notable, at 45%. Regarding the decision to migrate, 51.1% followed their children's suggestions, 33.0% decided independently, and 16.0% moved following recommendations from spouses or children. Data surveys reveal that their issues mainly revolve around "difficulties in communication due to relocation", "alienation between generations " and "inadaptability to new customs ". The reasons for these problems are closely related to the low ubiquity of today's digital technology, the fact that these groups do not have mature social relationships in new cities, are difficult to contact and have difficulty in accurately and proactively discovering and describing their needs.

Based on the meta-analysis, we invited 9 older adults (female 4, male 5, mean age= 68.33 SD=5.715) who had migrated to Beijing from other provinces in China with their children. With each participant, we conducted a simple 20-minute semi-structured interview. These interviews, conducted from various perspectives, delved into their current living conditions, migration experiences, family member status, social life, daily routines, etc. The interviews took place in a relaxed setting, with researchers facilitating the process to minimize the interaction burden on the older participants. This older-focused and gradual approach was carefully planned to prevent any physiological or psychological distress.

---

[1] 2021 "Old drifters", the physical and mental drift of the double predicament how to crack https://news.cctv.com/2021/01/14/ARTIQGT5cVLv6mrSFelfPVv5210114.shtml

Considering the positive role of Cognitive behavioral therapy [20], we prepared several 3D scene materials in advance to assist with the interviews. We used real-scene scanning software such as 3D Scanner to conduct model scanning, processing, and VR development of some indoor and outdoor environments such as "hutongs" and residential areas in Beijing. This was used as interview materials provided to participants during the interviews to help them understand VR technology and 3D scene scanning technology. This also facilitated further discussions with them about the potential of combining VR technology with reminiscence.

## 3. Insights

### 3.1. Current Living Conditions

In interviews, participants generally believed that they did not have too many worries about material life, but they continued to repeat the same numb life and work every day. They frequently mentioned the significant role of their cultural identity from their hometowns in their lives post-migration. This nostalgia often left them caught between the cultural identities of their hometown and their new residence. Many older internal migrants have encountered a certain degree of social problems in the new environment. P6 spoke of preferring solitude after moving to the new city, taking up photography post-retirement, and believing in the quality, not quantity, of friendships. *"I have only a few close friends. As you know, perhaps women's minds are also thoughtful, unlike most men who can easily get along with others. My acquaintances are mostly former colleagues, but I wouldn't call them friends."* For those who are introverted or haven't established stable social connections in parks, the desire to find new friends or forms of entertainment through technological assistance is particularly strong. P8 said, *"I don't have many friends here, I just wander around in the park without really meeting anyone. I'm not keen on participating in activities, just watching from the sidelines is good enough for me."* But when we asked him if he was interested in having entertainment or social activities through the VR platform we built, he expressed his willingness to participate.

### 3.2. About Memories

When reflecting on their hometown memories, approximately one-third of the participants expressed missing their local friends and family. Further investigation revealed a significant emphasis on the yearning for parents, beyond just peers and colleagues, among older internal migrants. This emotional connection to parents is crucial for providing support, significantly influencing their memories of home. One participant noted, *"I used to return to my hometown often when my parents were there. Now, with my parents gone, even though I have siblings, my visits have become less frequent, nearly a decade apart..."* The experience of migration introduces physical distance between individuals, fostering a sense of regret and helplessness among older internal migrants. This issue has become a prominent focus of HCI technology in enhancing the social connections of older adults [21].

On the other hand, around the location, the concept of "hometown" plays a significant role in the lives of immigrants [22]. Surprisingly, during the interviews, most older internal migrants expressed that they actually do not remember and miss the environment of their hometown as much as the public spaces in their hometown. Compared with the houses that remind them of their lonely lives, they prefer open spaces such as doorsteps or yards where they can greet and chat with others. The nostalgia is higher. In their hometowns, many older adults lived alone or with their partners. During the interviews, p1 revealed in words how bored and lonely she felt when living alone in an empty home. She said, *"Back in my hometown, I would just lie or sit upstairs all day. What's the point? I lived in a big courtyard,*

*and I could see fields with fruits and vegetables growing. I could visit them every day, which is different from this city."* P8 also mentioned the place he missed the most, saying, *"I used to lie in the front yard sunbathing, one day after another, without much to do. In the afternoon, when I heard a group of people coming from afar, making noise and chatting, I would pack up and go to the courtyard to play cards with them. I often miss those days now."* P8 described the mixed effects of recalling people, events, and scenes from the past. The courtyard in his hometown and the voices of his friends provided him with a sense of comfort, allowing him to piece together these mixed fragments and experience a moment of mental relaxation amidst vague memories.

Meanwhile, some participants spontaneously mentioned the enduring impact of their childhood in their hometown on their lives. The presence of their hometown is like a tangible memory probe, carrying with it the events, social relationships, and emotional responses that occurred there [23]. P7 said, *"I really miss the carefree life of my youth when I could just take a bath in the river in front of my home and catch fish there. But now, that river has dried up."* P7 described the contradiction between his cherished memories and the present reality. This change in the real environment makes childhood memories less evidence and fewer clues to trigger more memories. *"The disappearance of the river still affects me. I wish I could see that river again..."* Recreating changing scenes can help people draw strength from their memories. After we presented the VR and 3D scene reconstruction materials, the participants felt very excited. P7 said, *"Sharing these memories with everyone, being able to see things that have disappeared, is beautiful and meaningful."* In the context of the uncertain new migrant environment, older adults become newcomers, and they need their old life to motivate them to face a new life. At the same time, it provides immersive scenes for older internal migrants to reconnect with their memories of the past and offers a potential channel.

### 3.3. About Immersive Scenarios and VR

Based on the interview results, most older adults gave a favorable evaluation of the potential role of immersive scenes, and are highly receptive to realistically restored three-dimensional memory scenes, which breaks the limitations of 2D photos and videos on mobile phones and transcends physical limitations. P7 stated, *"It's like walking back in time."* However, some participants also showed a high level of vigilance, expressing concerns about data privacy and property safety, as well as doubts about the feasibility of the technology. Some older adults felt that due to their age, they lacked the confidence to try new technologies, speculating that they might not fully understand and use VR equipment. Older adults were hesitant to try VR devices due to their unfamiliar appearance and complete visual enclosure.

Furthermore, through the initial exploration of user needs mentioned above, we noticed that most participants prefer VR scenes that are highly restored and realistic. VR is a promising and ecologically valid technique to simulate physical environments and examine landscape preferences and restorativeness. A substantial amount of evidence suggests that under certain conditions, natural simulations can support processes that promote health. More than 100 experimental reports indicate that pictures, videos, or immersive virtual environments with natural elements can enhance mood, improve executive cognitive functions, promote physiological stress recovery, or alleviate pain [24, 25]. This approach fully leverages the capabilities of metaverse concepts and technologies.

### 3.4. About Resilience

At a time when they should be enjoying their old age, older internal migrants need to rebuild a new life in a new environment. Communication barriers from varied cultural backgrounds and lifestyle challenges due to different routines are distinct difficulties older internal migrants face, compared to other older adults. Seemingly trivial matters of daily life can slowly erode their self-esteem and confidence, like snowflakes. However, through brief interviews, we found that they could trigger energy in a moment of reminiscence, regaining the happiness and comfort of the past. This is not just about amplifying positive past emotions but using reminiscence to offset new life's emotional challenges, highlighting memory's restorative power as tacit knowledge. As older internal migrants, they may become lost in the difficulties of a new life, forgetting the meaning of their past lives. They need to be reminded that they also have the right and ability to live comfortably.

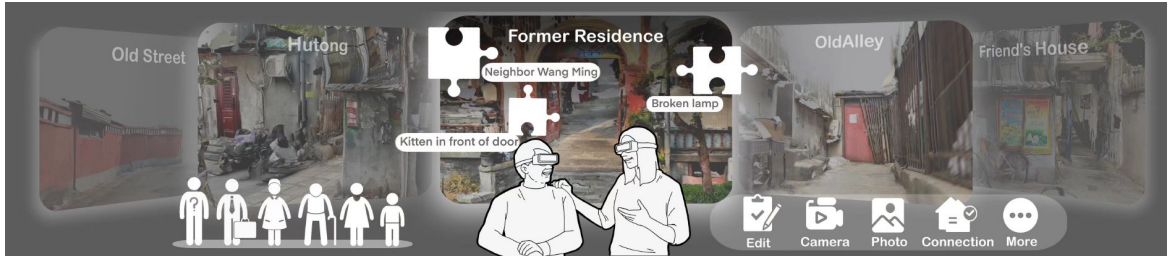

**Figure 2:** Enables simultaneous participation of older internal migrants and researchers in immersive scenarios for discussing, disclosing, defining, or editing memories.

From interviews with older internal migrants, we have confirmed the effective impact of memories in fostering positive self-emotions. Regrettably, the negative effects of aging on episodic memory recall extend to diminished self-awareness [26, 27]. Therefore, efforts to support self-defining memories can be beneficial to the sense of self and confidence, by tapping into the sensorial richness of episodic memories underpinning them [28]. In future participatory design, we plan to incorporate detailed disclosures from older adults, providing them with more tools and materials to strengthen their memories.

## 4. Conceptual Design Combining the Metaverse and Memory

We propose a metaverse design concept that promotes multi-party participation. This concept allows for experimentation with real scene models as core elements of immersive situational maps, aimed at facilitating reminiscence. Such an approach could enable older adults to more clearly share their memories, offering insights beyond those typical of conventional face-to-face interviews. In the early development phase, designers will guide older adults in enhancing the recreation of their memory worlds. Additionally, we advocate for incorporating insights from previous work on multimodal situational maps in reminiscence therapy into this design concept, such as integrating contextual maps with paper drawings [29], participatory mapping [30], and combinations of sound, smell, and images in contextual maps [31].

We propose that the concept design for metaverse reminiscence scenes should encompass four key elements: scenes, events, characters, and props (Figure 2). We found that changes in the living environment are the fundamental way to explore the life difficulties of older internal migrants. The parts involving reminiscence and scene recreation serve as powerful tools to manifest the impact of these changes [32]. For older internal migrants, changes in their living places have a significant impact. At the scene level, we assist users or other family members and friends to use mobile devices with depth cameras to scan the scenes of past and present memories or activities of the older internal migrants, and the designers will process and upload them. For unclear or potentially irrecoverable memory segments, as well as events and characters emerging from new social engagements encountered by older internal migrants, prior studies indicate that memory reinforcement can be achieved with methods like marking and annotation [33], and the use of virtual artifacts [34]. These techniques may aid older adults in memory sharing and social interaction. Consequently, we propose integrating props into the metaverse design. Designers and researchers can provide simple stickers, models, and other materials in VR to assist older adults in disclosing their views and evaluations of past or current events and characters, reflecting on their past ways of facing life difficulties and stimulating emotional resilience. Additionally, we could gather more design suggestions to enhance the initiation of upcoming research effectively.

To develop a VR system accessible to older adults with limited tech experience, as well as stakeholders, designers, and researchers, we have identified several key criteria. These include multi-user support, quick scene model uploading, the integration of various props like images and text, and customizable basic interactive features. Given these requirements, we find the 3Dscanner mobile app to be highly suitable for scene scanning. Similarly, VR social platforms such as VRchat and VRreborn emerge as ideal for facilitating scene interactions.

## 5. Discussion and Future Work

In this research, we identified the significant challenges faced by a large population of older internal migrants, caught between new, unfamiliar surroundings and a past that is out of reach. During our discussions, these individuals revealed a deep sense of nostalgia and a desire to reminisce. This suggests that memories play a crucial role in their resilience. Additionally, the positive response from these migrants towards VR technology demonstrates the effectiveness of immersive metaverse environments in facilitating their recollection of past experiences. Our research highlights the value of designing metaverse interventions that bolster emotional resilience [35] and self-esteem [36] among older internal migrant populations. We aim to delve deeper into the prerequisites for this concept's realization, developing a metaverse reminiscence environment enriched with older adults' detailed experiences. This initiative will enable older internal migrants and HCI researchers to engage in collaborative design, fostering mutual support. Our goal is to uncover and address the social challenges encountered by these migrants. Furthermore, we plan to refine the strategies HCI researchers employ for documentation and guidance, striving to craft a metaverse realm that significantly improves the living conditions of this demographic.

## 6. Conclusion

In this study, we began by examining the resilience of older internal migrants. Our findings reveal the beneficial effects of the metaverse concept and memory on their lives after migration. We propose the use of metaverse technology to enable older internal migrants to rebuild their self-identity from a personal viewpoint. By incorporating immersive real-world scenarios with characters, events and props, we aim to innovate memory recall methods, encouraging sharing and clearer memory articulation. This technique also enhances designers' and researchers' understanding of user expressions, offering new solutions and methods for future challenges.

## Acknowledgments

This work was supported by the Talent Fund of Beijing Jiaotong University(Grant No. 2023XKRCW002).